\documentclass[fleqn,10pt]{wlscirep}
\title{2D layered transport properties from topological insulator Bi$_2$Se$_3$ single crystals and micro flakes}
\author[1]{Olivio Chiatti}
\author[1]{Christian Riha}
\author[1]{Dominic Lawrenz}
\author[1]{Marco Busch}
\author[1]{Srujana Dusari}
\author[2]{Jaime S\'anchez-Barriga}
\author[3]{Anna Mogilatenko}
\author[4]{Lada V. Yashina}
\author[2]{Sergio Valencia}
\author[2]{Akin A. \"Unal}
\author[2]{Oliver Rader}
\author[1,*]{Saskia F. Fischer}
\affil[1]{Novel Materials Group, Humboldt-Universit\"at zu Berlin, Newtonstra{\ss}e 15, 12489 Berlin, Germany}
\affil[2]{Helmholtz-Zentrum-Berlin f\"ur Materialien und Energie, Albert-Einstein-Stra{\ss}e 15, 12489 Berlin, Germany}
\affil[3]{Ferdinand-Braun-Institut, Leibniz-Institut f\"ur H\"ochstfrequenztechnik, Gustav-Kirchhoff-Stra{\ss}e 4, 12489 Berlin, Germany}
\affil[4]{Department of Chemistry, Moscow State University, Leninskie Gory 1/3, 119991 Moscow, Russia}
\affil[*]{Correspondence and requests should be addressed to S.F.F. (email: sfischer@physik.hu-berlin.de)}
\begin{abstract}
\textbf{Low-field magnetotransport measurements of topological insulators such as Bi$_2$Se$_3$ are important for revealing the nature of topological surface states by quantum corrections to the conductivity, such as weak-antilocalization. Recently, a rich variety of high-field magnetotransport properties in the regime of high electron densities ($\sim10^{19}$ cm$^{-3}$) were reported, which can be related to additional two-dimensional layered conductivity, hampering the identification of the topological surface states. Here, we report that quantum corrections to the electronic conduction are dominated by the surface states for a semiconducting case, which can be analyzed by the Hikami-Larkin-Nagaoka model for two coupled surfaces in the case of strong spin-orbit interaction. However, in the metallic-like case this analysis fails and additional two-dimensional contributions need to be accounted for. Shubnikov-de Haas oscillations and quantized Hall resistance prove as strong indications for the two-dimensional layered metallic behavior. Temperature-dependent magnetotransport properties of high-quality Bi$_2$Se$_3$ single crystalline exfoliated macro and micro flakes are combined with high resolution transmission electron microscopy and energy-dispersive x-ray spectroscopy, confirming the structure and stoichiometry. Angle-resolved photoemission spectroscopy proves a single-Dirac-cone surface state and a well-defined bulk band gap in topological insulating state. Spatially resolved core-level photoelectron microscopy demonstrates the surface stability.}
\end{abstract}
\begin{document}
\flushbottom
\maketitle
\thispagestyle{empty}
\section*{Introduction}

A tremendous interest in the electronic properties of topological insulators (TIs),\cite{kon07,kan05} such as Bi$_2$Se$_3$, stems from the fact that they have robust topological surface states (TSS).\cite{kan05,fu07,moore07,has10} The existence of TSS in Bi$_2$Se$_3$ was observed by angle-resolved photoemission spectroscopy and scanning tunneling spectroscopy.\cite{hsi08,xia09,che09,alp10,pan11} While TSS are predicted to have peculiar properties of great interest for future electronic devices in spintronics and quantum computation,\cite{Fu09,Akh09,has11,sem12,wang12,sch12,che12} the direct access in transport experiments still remains to be demonstrated. The defect chemistry in Bi$_2$Se$_3$ is dominated by charged selenium vacancies, which act as electron donors, and increase the conductivity of bulk states dramatically.\cite{Hor} However, it was shown that the TSS coexist with the Se vacancies and that in ARPES measurements a single Dirac cone can be observed, even though Se vacancies exist.\cite{Felser} In particular, recent reports\cite{Analytis,Petrushevsky-2012-prb,caoPRL,Yan} on highly doped Bi$_2$Se$_3$ additionally show indications of two-dimensional (2D) layered transport, which hampers the unambiguous identification of the TSS. For high electron densities ($\sim10^{19}$ cm$^{-3}$) a rich variety of magnetoresistance phenomena in high-magnetic fields were detected, such as Shubnikov-de Haas (SdH) oscillations and quantum Hall resistances, and attributed to the behavior of stacked 2D electron systems. However, their interpretations are controversial.\cite{Analytis,Petrushevsky-2012-prb,caoPRL}

In this work, we report on TSS, which were probed by angle-resolved photoemission spectroscopy (ARPES), and significant 2D layered transport properties from both, the high- and low-field magnetoresistance of a high quality topological insulator Bi$_2$Se$_3$ single crystal. By exfoliation macro and micro flakes were prepared to investigate bulk and surface contributions (see Sec. Methods). Comprehensive combined structural, electronic and low-temperature magnetotransport investigations show: Quantum corrections to the electronic conduction are dominated by the TSS carriers in the semiconducting regime, but in the metallic-like regime additional 2D layers contribute. In low-magnetic fields a weak-antilocalization (WAL) cusp in the conductivity exists in both cases, and we discuss the analysis employing the Hikami-Larkin-Nagaoka (HLN) model.\cite{hik80}. As support for the 2D layered transport in the metallic-like case, we find SdH oscillations in the longitudinal bulk conductivity and quantization of the transversal (Hall) resistivity in high-magnetic fields.

\section*{Results}

\subsection*{Structural characterization}

\begin{figure}[h!]
\begin{center}
\includegraphics[width=0.6\columnwidth]{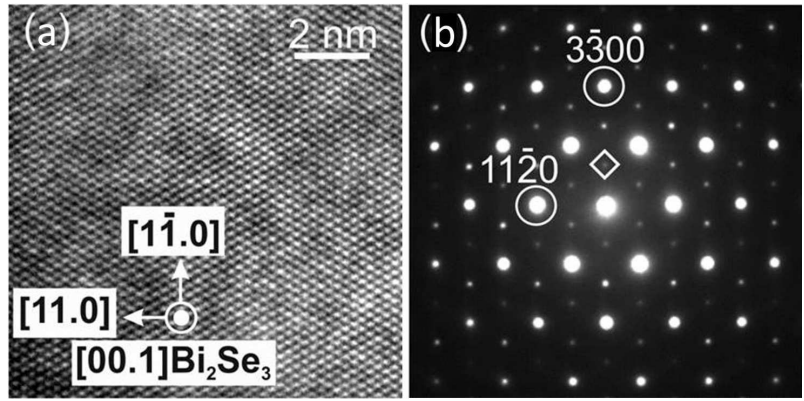}
\caption{(a) HRTEM image of the flake region of about 40 nm in thickness, viewed in the direction of the surface normal. (b) SAED pattern viewed in [00.1] direction of the Bi$_2$Se$_3$. The diamond indicates the $\{1\bar{1}00\}$ reflection.}
\label{fig1}
\end{center}
\end{figure}

A high-quality single crystal of nominally undoped Bi$_2$Se$_3$ was grown by the Bridgman technique.\cite{sht09} The typical high-resolution transmission electron microscopy (HRTEM) image in Fig.\ref{fig1}(a) from a homogeneously thick region of the flake, viewed in the direction perpendicular to the surface, shows a 2D arrangement of lattice fringes with 6-fold symmetry and a lattice spacing of $\sim$0.21 nm. This proves that the surface normal of the exfoliated flake is parallel to the [00.1] Bi$_2$Se$_3$ zone axis. 

The selected area electron diffraction (SAED) pattern shown in Fig.~\ref{fig1}(b), obtained from a large region of the flake, reveals the single crystal nature of the sample. Interestingly, the SAED pattern shows the presence of weak superstructure reflections, indicating possible ordering effects appearing at the $\{1\bar{1}00\}$ planes. Similar diffraction patterns were reported for Bi$_2$Se$_3$ and Bi$_2$Te$_3$ nanoribbons and nanoplates.\cite{kon10,xiu11,wan03} For well-ordered Bi$_2$Se$_3$ single crystals the structure factor of these additional reflections is zero. Furthermore, these reflections cannot appear in the [00.1] Bi$_2$Se$_3$ zone axis due to multiple scattering. According to the positions of the superstructure reflections, in the (0001) plane the Bi$_2$Se$_3$ supercell is three times larger than the Bi$_2$Se$_3$ unit cell, which is built by $\mathbf{a}\times\mathbf{b}$ unit vectors. For example, a similar superstructure has been previously assumed in Cu$_x$Bi$_2$Se$_3$ single crystals, which raises the question if a similar effect can be caused by bulk dopants.\cite{han11} Indeed, such ordering effects can be caused by a strictly periodic arrangement of point defects at the corresponding atomic planes (vacancies, substitutional atoms, excess of Bi or Se). 

A second possible reason for the appearance of the $\{1\bar{1}00\}$ reflections can be a certain structural stacking disorder along the $c$-axis direction. For example, diffraction pattern simulations we carried out for a Bi$_2$Se$_3$ unit cell with one missing quintuple layer show the appearence of these reflections. This might appear at the Bi$_2$Se$_3$ surface and can become especially pronounced for thin Bi$_2$Se$_3$ layers and nanostructures, where the volume fraction of disordered surface regions is larger. Our HRTEM analysis did not reveal any visible superstructure in the $[00.1]$ zone axis of Bi$_2$Se$_3$ (Fig.~\ref{fig1}(a)), which further supports this explanation. Therefore, we attribute this diffraction effect to a contribution by the surface. A detailed electron diffraction analysis of bulk Bi$_2$Se$_3$ single crystals and flakes of different thicknesses is required to unambiguously identify the origin.

\subsection*{Electronic structure}

\begin{figure}[h!]
\includegraphics[width=1\textwidth]{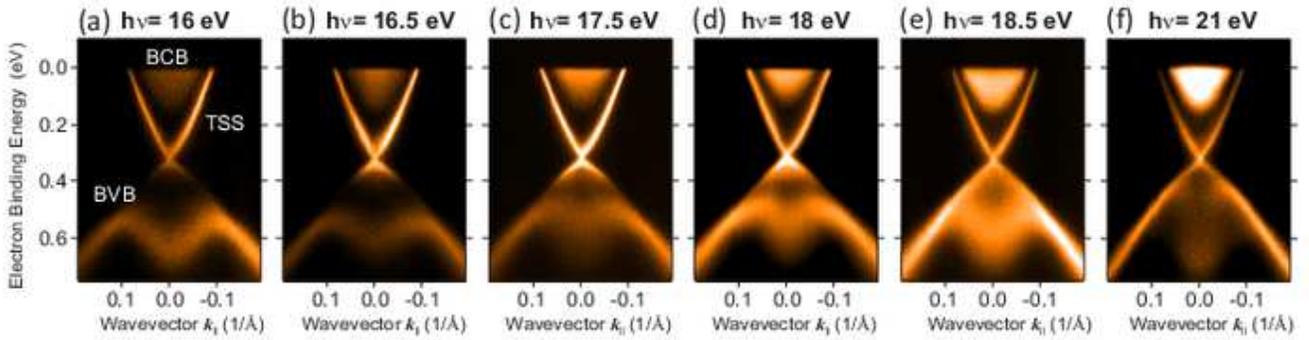}
\caption{$k_{\mathrm{z}}$-dependence of the electronic structure of the Bi$_2$Se$_3$ bulk single crystal before mechanical exfoliation. Each panel shows high resolution ARPES $E(k_\parallel)$ dispersions measured at 12 K and at photon energies $h\nu$ of (a) 16~eV, (b) 16.5~eV, (c) 17.5~eV, (d) 18~eV, (e) 18.5~eV and (f) 21~eV across the first bulk Brillouin zone. In panel (a), the topological surface state (TSS), the bulk conduction band (BCB) and the bulk valence band (BVB) are labelled. The surface state does not show dispersion with photon energy, while bulk states exhibit a clear dependence.}
\label{fig2}
\end{figure}

To verify the characteristics of the surface states of the Bi$_2$Se$_3$ bulk single crystal, high resolution ARPES experiments were performed at different photon energies. The original $(0001)$ surface was characterized to determine its electronic structure and the Fermi level position or intrinsic doping level of the samples. Fig.~\ref{fig2} shows high-resolution ARPES dispersions of the TSS, bulk conduction band (BCB) and bulk valence band (BVB) states, measured at different photon energies and as a function of the electron wave vector $k_\parallel$ parallel to the surface. A gapless Dirac cone representing the TSS with a Dirac point located at a binding energy of $\sim$0.35 eV is clearly observed. The binding energy position of the BCB crossing the Fermi level indicates that the crystals are intrinsically $n$-type, in agreement with our Hall measurements. At binding energies higher than the Dirac point, the lower half of the Dirac cone overlaps with the BVB. In ARPES, the photon energy selects the component of the electron wave vector $k_\mathrm{z}$ perpendicular to the surface. Since the lattice constant of Bi$_2$Se$_3$ is very large along the z direction ($c\!=\!28.64$ \AA), the size of the first bulk Brillouin zone is very small ($\sim$0.5 \AA$^{-1}$). Therefore, Figs. \ref{fig2}(a)-\ref{fig2}(f) show a range of low photon energies between 16 to 21~eV, where we practically cross the complete Bi$_2$Se$_3$ first bulk Brillouin zone enhancing the sensitivity to the out-of-plane dispersion of the bulk bands. We note that the maximum ARPES intensity changes with the photon energy as well due to the $k_{\mathrm{z}}$-dependence of the photoemission transitions. 

Unlike the BCB or the M-shaped dispersion of the BVB, the TSS has no $k_{\mathrm{z}}$-dependence, a fact which confirms its 2D nature. Its dispersion remains very clearly the same when varying the photon energy, as can be seen in Figs. \ref{fig2}(a)-\ref{fig2}(f). In contrast, the BVB maximum is reached around at $h\nu\!=\!18$ eV ($k_{\mathrm{z}}\!=\!2.65$ \AA$^{-1}$) and the BCB minimum near 21~eV ($k_{\mathrm{z}}\!=\!2.8$ \AA$^{-1}$) at binding energies of $\sim$0.452 eV and $\sim$0.154 eV, respectively. This is consistent with a bulk band gap of about $\sim$0.3~eV. These facts unambiguously identify the existence of both a single-Dirac-cone surface state and a well-defined bulk band gap in our samples, two of the most important attributes of the topological insulator Bi$_2$Se$_3$. The data in Fig.~\ref{fig2} allow us to estimate the bulk carrier concentration $n_\mathrm{3D}$ from the size of the bulk Fermi wave vector $k_{\rm F,3D}$ according to $n_\mathrm{3D}\!=\!k_{\rm F,3D}^3/(3\pi^2)$. The limited accuracy of $k_{\rm F, 3D}\!=\!(0.064\pm0.01)$ \AA$^{-1}$ yields $n_\mathrm{3D}\!=\!(8.8\pm0.4)\cdot10^{18}$ cm$^{-3}$.

\subsection*{Core-level spectroscopy}

\begin{figure}[t!]
\begin{center}
\includegraphics[width=0.49\columnwidth]{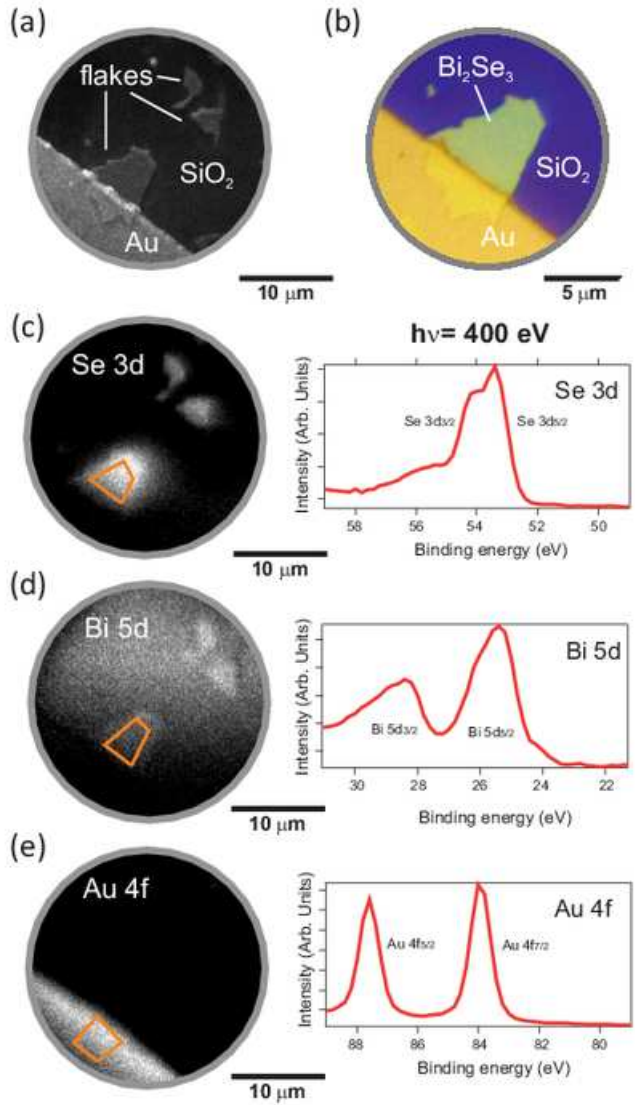}
\caption{PEEM characterization of a contacted Bi$_2$Se$_3$ flake. (a) Overview image with vacuum UV light of the flake and Au contact area. (b) Corresponding confocal microscopy image of the same flake revealing thickness homogeneity. (c), (d) and (e): Core-level spectromicroscopy of the flake area and of the Au contact area. On the left, spatially resolved images taken at the kinetic energies of the (c) Se 3d$_{5/2}$, (d) Bi 5d$_{5/2}$ and (e) Au 4f$_{5/2}$ core-levels. On the right, the corresponding core-level spectra are shown. The spectra are extracted from the small areas indicated in each PEEM image on the left with a kinetic energy resolution of $\sim$0.2~eV. The flakes are stable to atmosphere and to the lithography process. The measurements were performed using soft x-rays of 400~eV photon energy and horizontal polarization. The PEEM field of view is 25 $\mu$m.}
\label{fig3}
\end{center}
\end{figure}

In order to investigate the chemistry of the micro flake surface after exfoliation and lithographic processing of electrical contacts for transport measurements, we performed spatially resolved core-level X-ray photoelectron emission microscopy (PEEM). It serves the following purposes: a test of stability of the micro flake surface in high-electric fields, a test of disturbance by charging of the SiO$_2$ substrate, and in particular, confirmation of the chemical composition against ambient atmosphere providing useful information about the oxidation state of the flakes. The experiments were performed at a photon energy of 400 eV and horizontal polarization. The kinetic energy resolution was $\sim$0.2 eV and the lateral resolution about 70~nm. Fig.~\ref{fig3}(a) shows an overview PEEM image of a Bi$_2$Se$_3$ micro flake with a Ti/Au contact acquired in vacuum with UV light. Fig.~\ref{fig3}(b) shows the corresponding confocal microscopy image, which zooms into the same flake and reveals its thickness homogeneity. The core-level spectromicroscopy results of the flake area and of the Au contact area are shown in Figs. \ref{fig3}(c)-\ref{fig3}(e). Spatially resolved images acquired at the kinetic energies of the Se 3d$_{5/2}$ (Fig.~\ref{fig3}(c)), Bi 5d$_{5/2}$ (Fig.~\ref{fig3}(d)) and Au 4f$_{5/2}$ (Fig.~\ref{fig3}(e)) core-levels are shown together with the corresponding core-level spectra. The spectra are extracted from the small areas indicated in each PEEM image. 

Our results demonstrate that the flakes and their composition are stable to ambient atmosphere and to the lithography process. However, we do note that the Se 3d core-levels show a small component separated by $\sim$2 eV to higher binding energy. The Bi 5d peaks exhibit two small additional components which are located $\sim$1~eV higher in binding energy with respect to the main peaks and are not well-resolved due to the experimental energy resolution. These findings indicate that a minor oxidation of the Bi$_2$Se$_3$ flake surface exists. Our estimations reveal that oxygen is adsorbed in less than 15\% of a quintuple layer. Moreover, this oxygen adsorbate layer was easily removed by moderate annealing under ultra-high vacuum (UHV) conditions at $\sim120-180\,^{\circ}{\rm C}$ during 5 minutes. 

Recently, we observed a similar surface reactivity by means of ARPES experiments in topological insulator bulk single crystals and films grown by molecular beam epitaxy.\cite{Yashina-ACSNano-2013} We found that the topological surface state in our samples is robust against the effects of surface reactions, which typically lead to doping and extra quantization effects in the ARPES dispersions due to band bending.\cite{Bianchi-NatComm-2011,King-PRL-2011,Bianchi-PRL-2011} By means of ARPES and core-level photoemission experiments at different pressures, we demonstrated that our Bi$_2$Se$_3$ bulk single crystals exhibit a negligible surface reactivity toward oxygen and water.\cite{Yashina-ACSNano-2013}

\subsection*{Temperature-dependent longitudinal resistances}

\begin{figure}[t!]
\begin{center}
\includegraphics[width=0.47\columnwidth]{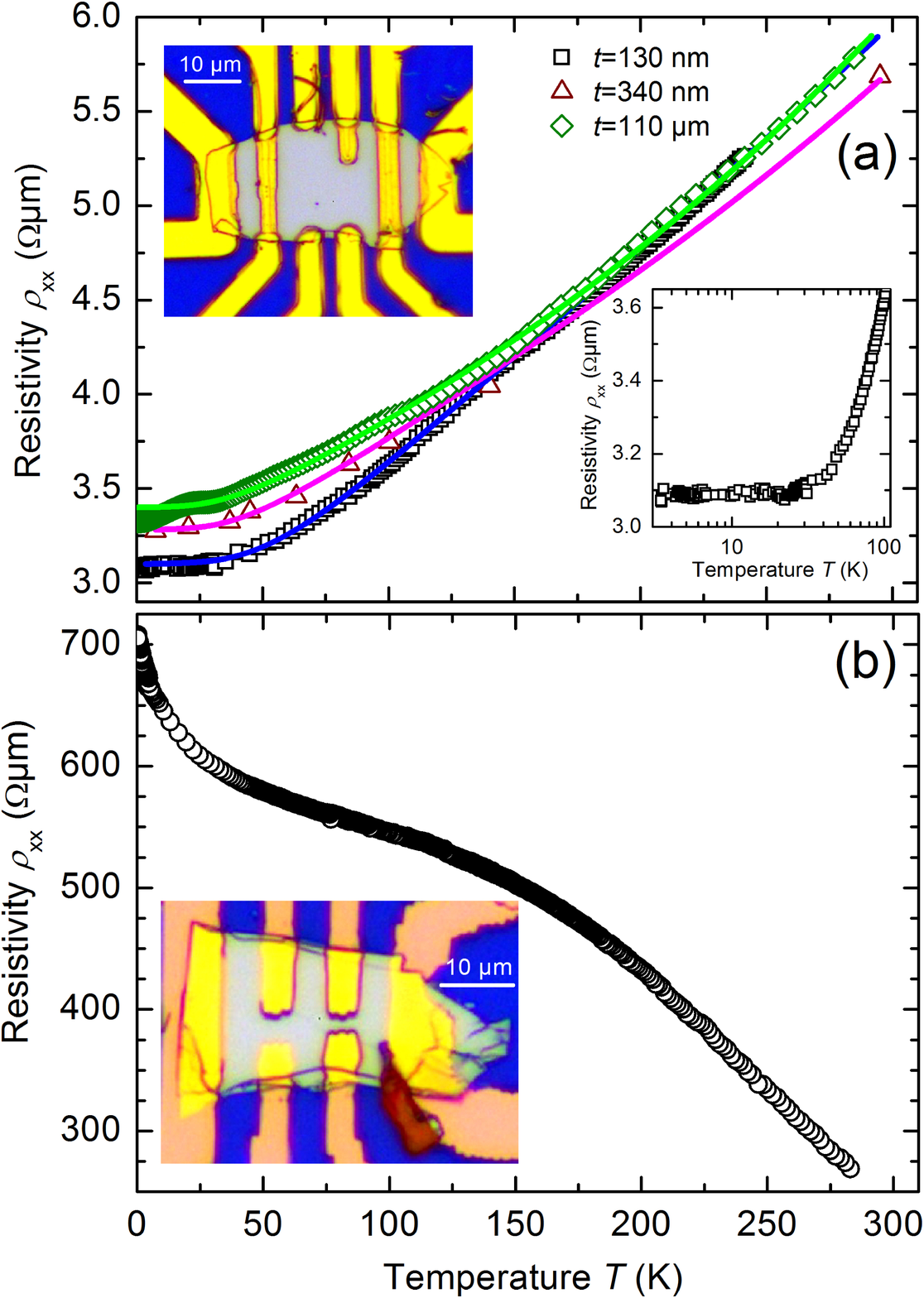}
\caption{(a) Temperature dependent resistivity $\rho_{\mathrm{xx}}$ of two Bi$_2$Se$_3$ micro flakes with thicknesses of $t\!=\!130$ nm (black squares) and 340 nm (brown triangles) and a Bi$_2$Se$_3$ macro flake with a thickness of $t\!=\!110$ $\mu$m (green diamonds). Insets in (a) show an optical microscope image of the Bi$_2$Se$_3$ micro flake with a thickness of $t\!=\!130$ nm with Ti/Au contacts, and the low temperature range of the measured resistivity $\rho_{\mathrm{xx}}$ of this micro flake (with a logarithmic temperature axis). Solid curves represent best Bloch-Gr\"uneisen fits. (b) Temperature dependent resistivity $\rho_{\mathrm{xx}}$ of a semiconducting Bi$_2$Se$_3$ micro flake with a thickness of $t\!=\!220$ nm. Inset in (b) shows the optical microscope image of this micro flake with Ti/Au contacts.}
\label{fig4}
\end{center}
\end{figure}

Bi$_2$Se$_3$ flakes were contacted in Hall bar geometries for transport measurements. Two examples are shown in the insets of Fig.~\ref{fig4}(a) and (b) by optical microscope images for thicknesses of $t\!=\!200$ nm (inset of Fig.~\ref{fig4}(a)) and $t\!=\!220$ nm (inset of Fig.~\ref{fig4}(b)), respectively. In the temperature-dependent longitudinal resistance we observe two cases: metallic-like behavior as shown Fig.~\ref{fig4}(a) and semiconducting behavior as shown Fig.~\ref{fig4}(b). In Fig.~\ref{fig4}(a) the metallic-like behavior of the resistivity $\rho_{\mathrm{xx}}$ as function of temperature from 4.2~K up to 290~K for two typical Bi$_2$Se$_3$ micro flakes with thicknesses of $t\!=\!130$ nm and 340 nm and for a macro flake with a thickness of $t\!=\!110\;\mu$m is shown. The resistivity $\rho_{\mathrm{xx}}$ remains practically constant up to 30 K (see right inset in Fig.~\ref{fig4}(a)), presumably due to a combination of surface states and static disorder scattering, as observed in the previous reports.\cite{kim11,ban12,che10,he12,he11} The residual resistivity ratio $\mathrm{RRR}\!=\!\rho_{\mathrm{xx}}(300~\mathrm{K})/\rho_{\mathrm{xx}}(2~\mathrm{K})\!=\!1.74$ for the Bi$_2$Se$_3$ macro flake indicates the high crystalline quality.\cite{Hyde} The RRR of the exfoliated micro flakes have similar values within $\pm20$\%, and for a 80~nm thin micro flake we found $\mathrm{RRR}\!=\!\rho_{\mathrm{xx}}(290~\mathrm{K})/\rho_{\mathrm{xx}}(0.3~\mathrm{K})\!=\!1.99$ indicating a decrease of bulk defects. In the temperature range between 100~K and room temperature Bloch-Gr\"uneisen fits can be performed. Such metallic-like behavior is in agreement with previous reports\cite{cao12} of exfoliated Bi$_2$Se$_3$ micro flakes with high electron densities of about $10^{19}$ cm$^{-3}$. However, as shown in Fig.~\ref{fig4}(b) semiconducting behavior in the temperature dependent resistivity $\rho_{\mathrm{xx}}$ can be also observed. This occurs if the fabrication and storage procedures of the micro flakes minimize any exposure to air (see Sec. Methods). Otherwise, metallic-like behavior is induced. Semiconducting behavior was also found for MBE grown Bi$_2$Se$_3$ thin films.\cite{Hirahara} 

\subsection*{Longitudinal and Hall conductivities}

While low-magnetic field conductivities were measured in order to determine the electron densities and mobilities, high-magnetic field conductivities show typical signatures of 2D transport, such as SdH oscillations and quantum Hall effect in macro flakes. The low-field conductivity $\sigma_\mathrm{xx}$ and the Hall conductivity $\sigma_\mathrm{xy}$ of the macro flake as a function of the perpendicular magnetic field $B$ at a temperature $T\!=\!4.1$ K are shown in Fig. S3 in the supplemental information. The conductivity curves were determined from the longitudinal resistivity $\rho_\mathrm{xx}$ and Hall resistivity $\rho_\mathrm{xy}$ as follows:\cite{Ando-2013-JPSJ} $\sigma_\mathrm{xx}\!=\!\rho_\mathrm{xx}/(\rho_\mathrm{xx}^2+\rho_\mathrm{xy}^2)$ and $\sigma_\mathrm{xy}\!=\!-\rho_\mathrm{xy}/(\rho_\mathrm{xx}^2+\rho_\mathrm{xy}^2)$. Both the $\sigma_\mathrm{xx}(B)$ and $\sigma_\mathrm{xy}(B)$ curves can be fitted within the Drude model and the results of the fits are in agreement with each other and yield an electron density $n_\mathrm{3D}\!=\!1.92\cdot10^{19}$~cm$^{-3}$ and mobility $\mu_\mathrm{3D}\!=\!970$ cm$^2$/(Vs). These values are consistent with those reported in a variety of highly-doped Bi$_2$Se$_3$ bulk crystals of similar resistivity.\cite{Hyde,Koehler,Petrushevsky-2012-prb,Ge-2015-ssc}

\begin{figure}[t!]
\centering
\includegraphics[width=0.5\columnwidth]{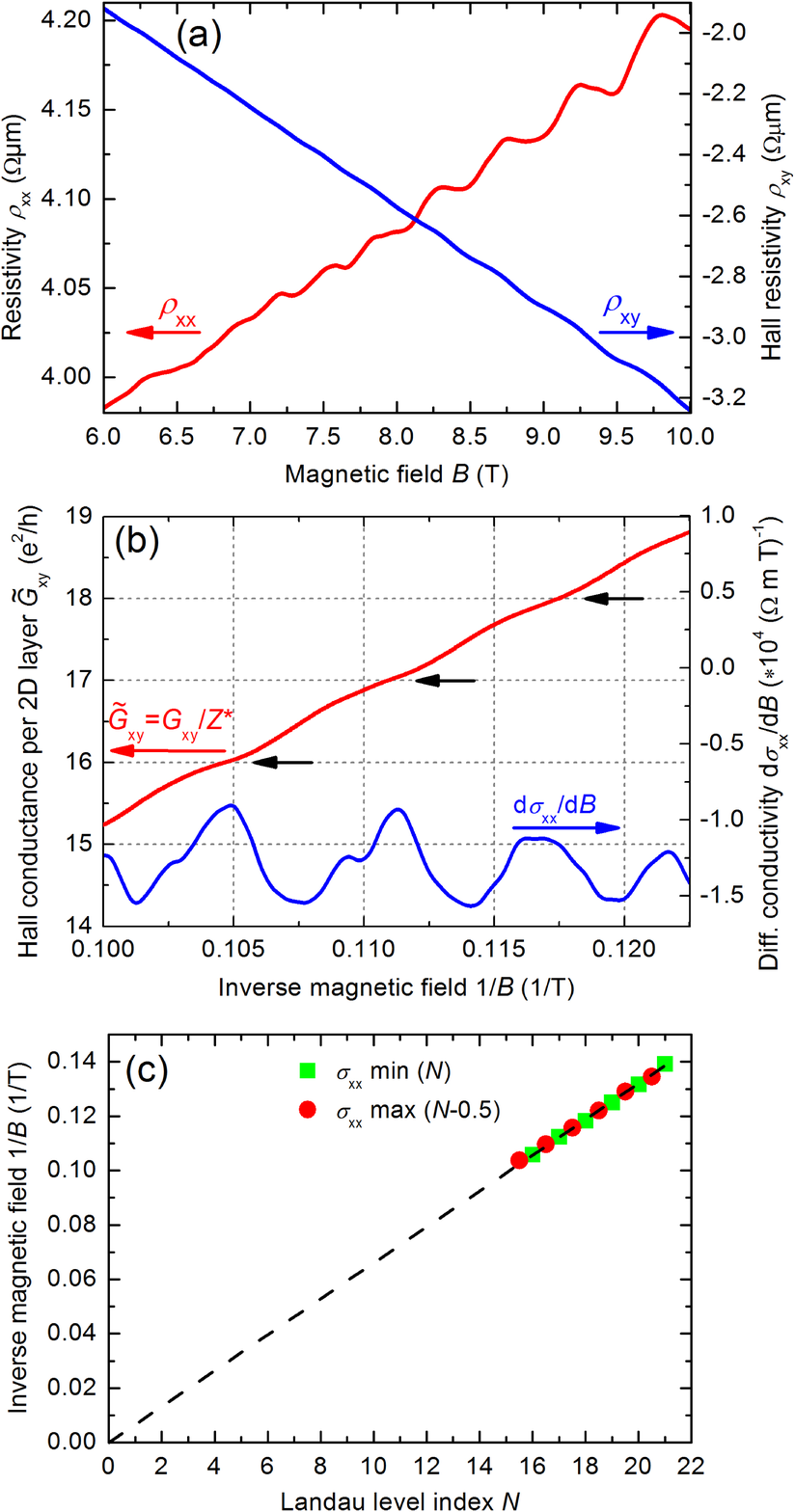}
\caption{(a) Resistivity $\rho_{\mathrm{xx}}$ (red curve, left axis) and Hall resistivity $\rho_{\mathrm{xy}}$ (blue curve, right axis) vs magnetic field $B$ of the Bi$_2$Se$_3$ macro flake with a thickness of $t\!=\!110$ $\mu$m at $T\!=\!0.3$ K. (b) Hall conductance per 2D layer $\widetilde{G}_\mathrm{xy}\!=\!G_\mathrm{xy}/Z^*$ in units of e$^2$/h (red curve, left axis), with measured conductance $G_\mathrm{xy}\!=\!1/R_\mathrm{xy}$ and $Z^*\!=\!57500$, and $\mathrm{d}\sigma_{\mathrm{xx}}/\mathrm{d} B$ (blue curve, right axis) vs inverse magnetic field $1/B$ at $T\!=\!0.3$ K. The black arrows indicate the QHE plateaux. (c) Landau level (LL) fan diagram at $T\!=\!0.3$ K. The $1/B$-positions of the minima and maxima of $\sigma_\mathrm{xx}(B)$ are shown as a function of the corresponding LL level indices $N$ and $N-0.5$, respectively. The dashed line represents a linear fit to the data, yielding a slope  $B_{\mathrm{f}}\!=\!151$ T and an intercept close to zero.}
\label{fig5}
\end{figure}

SdH oscillations in the resistivity $\rho_\mathrm{xx}$ and the onset of quantum Hall effect (QHE) plateaux in the Hall resistivity $\rho_\mathrm{xy}$ can be observed at high-magnetic fields. Fig.~\ref{fig5}(a) shows $\rho_\mathrm{xx}$ (red curve) and $\rho_\mathrm{xy}$ (blue curve) of the macro flake with a thickness of $t\!=\!110$ $\mu$m as a function of the perpendicular magnetic field $B$ at a temperature of $T\!=\!0.3$ K. The slope of $\rho_\mathrm{xy}$ yields an electron density $n_\mathrm{3D}\!=\!1.84\cdot10^{19}$~cm$^{-3}$ and the onset field $B\!\approx\!7.5$ T yields a mobility $\mu_\mathrm{3D}\!\approx\!1300$ cm$^2$/(Vs), both values in fair agreement with the results of the low-field transport. For a comparison with the values from ARPES measurements, one has to consider the non-spherical Fermi surface which plays a role in 3D bulk measurements.\cite{Ando-2013-JPSJ} This shows that for micron thick crystals both the low-field and the high-field magnetotransport are dominated by carriers from the bulk. However, low-temperature transport measurements with a magnetic field parallel to the current show no clear signs of SdH oscillations or QHE plateaux. This is consistent with 2D transport, instead of 3D transport, and has been reported previously in $n$-type Bi$_2$Se$_3$.\cite{caoPRL,cao12} Therefore, 2D layered transport plays an important role. 

In Ref. \cite{caoPRL} the steps in Hall conductance $\Delta G_\mathrm{xy}\!=\!\Delta(1/R_\mathrm{xy})$ were found to scale with the sample thickness and yield a conductance of $\sim$e$^2$/h per QL. A similar scaling was also found in Fe-doped Bi$_2$Se$_3$ bulk samples,\cite{Ge-2015-ssc} where transport by TSS can be excluded. From this it is concluded that the bulk transport occurs over a stack of 2D layers. Fig.~\ref{fig5}(b) shows such analysis performed on our high-field $R_\mathrm{xy}(B)$ data. We find a scaling of $\Delta G_\mathrm{xy}$ with the thickness and define $Z^*\!=\!\Delta G_\mathrm{xy}/(\mathrm{e}^2/\mathrm{h})$ as the number of 2D layers contributing to the transport, with $Z^*\!\approx\!0.5\times$ the number of QLs. Then $\widetilde{G}_\mathrm{xy}\!=\!G_\mathrm{xy}/Z^*\!=\!N\mathrm{e}^2/\mathrm{h}$ is the conductance per 2D layer, where $N$ is the Landau level index, and is shown in Fig.~\ref{fig5}(b). Therefore, a conductance of $\sim$e$^2$/h can be associated with an effective thickness of one 2D layer of about 2 QLs thickness, i.e. $\sim2$~nm. Note that this is similar the length scale of the extention of the volume unit cell of Bi$_2$Se$_3$ ($c$-axis, 2.86 \AA).\cite{Nakajima} The present results suggest that transport in macro flakes is consistent with a metallic-like 2D layered transport. Unfortunately, in our measurement setup for micro flakes the high-magnetic field ($>\!5$~T) magnetoresistance was not well enough resolved due to non-ideal contacts and time-varying thermovoltages. These effects may mask possible 2D transport signatures. 

The resistivity curves were used to determine the conductivity $\sigma_\mathrm{xx}$ and Hall conductivity $\sigma_\mathrm{xy}$ at high fields. Fig.~\ref{fig5}(c) shows the Landau level (LL) fan diagram determined from the measurements in Fig.~\ref{fig5}(a). The $1/B$-positions of the minima of the $\sigma_\mathrm{xx}$ curve are shown as a function of the corresponding LL level indices $N$. The LL indices have been attributed to the $\sigma_\mathrm{xx}$ minima as in Fig.~\ref{fig5}(b). Taking into account that the LL index of a maximum in $\sigma_\mathrm{xx}$ can be written as $N-0.5$,\cite{Ando-2013-JPSJ} the $1/B$-positions of the minima and the maxima collapse onto the same line. The slope of the line is the SdH frequency $B_{\mathrm{f}}\!=\!151$ T and it yields a Fermi wavevector $k_{\mathrm{F}}\!=\!0.0677$ \AA$^{-1}$, which is in fair agreement with $k_{\mathrm{F,3D}}$ from our ARPES results (see Fig.~\ref{fig2}), and does not agree with that of $k_{\mathrm{F,TSS}}\!=\!0.086$ \AA$^{-1}$. The intercept on the $x$-axis of the line yields the phase-factor $\beta$, which indicates whether fermions ($\beta\!=\!0$) or Dirac fermions ($\beta\!=\!0.5$) are responsible for the transport.\cite{Ando-2013-JPSJ} Within the experimental error the distinction between an intercept of 0 or 0.5 cannot be drawn unambiguously from the fan diagram. For this, higher magnetic fields such as 30 T are required. However, within the present high-field data the linear fit strongly suggests that the main contribution comes from the bulk. Conclusively, our present high-field results support a major contribution by 2D layered transport, additionally to the existence of TSS as proven by ARPES. 

\subsection*{Weak-antilocalization effects} 

\begin{figure*}[t!]
\centering
\includegraphics[width=0.8\columnwidth]{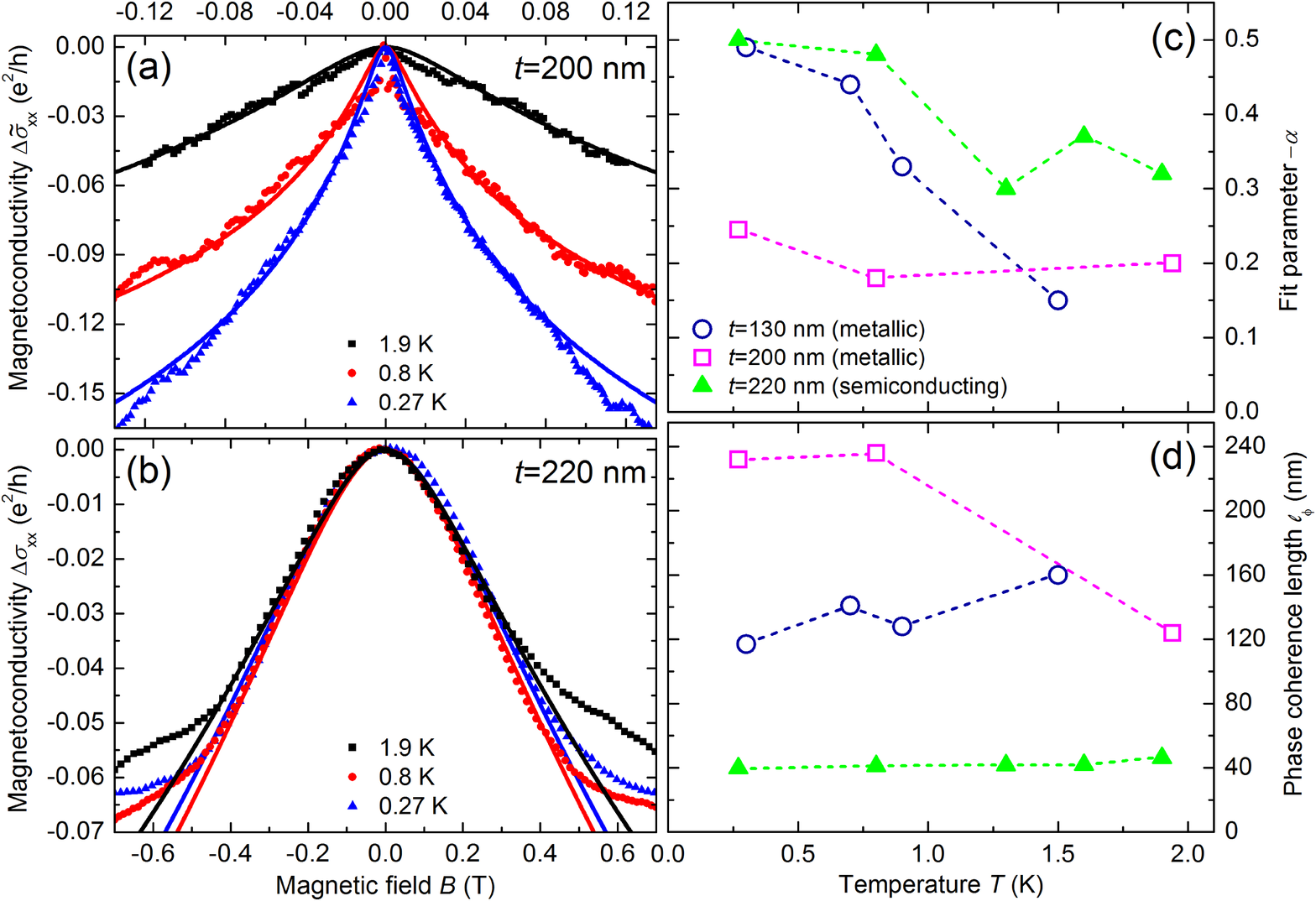}
\caption{Magnetoconductivity $\Delta\widetilde{\sigma}_\mathrm{xx}\!=\!\Delta\sigma_\mathrm{xx}/Z^*$ (metallic-like case) and $\Delta\sigma_\mathrm{xx}$ (semiconducting case) in the unit (e$^2$/h) vs magnetic field $B$ at $T\!=\!0.27$ K (blue triangles), 0.8~K (red circles) and 1.9~K (black squares) of Bi$_2$Se$_3$ micro flakes with thicknesses of $t\!=\!200$ nm (a) and $t\!=\!220$ nm (b). Curves present best HLN fits (for details see text). The fit \mbox{parameter} $\alpha$ and the phase coherence length $\ell_\phi$ of the HLN fits vs temperature $T$ are shown in (c) and (d), respectively, for Bi$_2$Se$_3$ micro flakes with thicknesses of $t\!=\!130$ nm (dark blue circles), 200 nm (magenta squares) and 220 nm (green triangles), respectively. For the HLN fits of the metallic-like micro flakes (with thicknesses $t\!=\!130$ nm and 200 nm) the magnetoconductivity $\Delta\widetilde{\sigma}_\mathrm{xx}\!=\!\Delta\sigma_\mathrm{xx}/Z^*$ with $Z^*\!=\!0.5\times$number of QLs was considered.}
\label{fig6}
\end{figure*}

As quantum correction to the classical magnetoresistance, the WAL effect is a signature of TSS originating from the Berry phase,\cite{Berry} which is associated with the helical states.\cite{Ando-2013-JPSJ} In the low-magnetic field longitudinal conductivity of exfoliated micro flakes we observe the typical WAL cusp. In macro flakes the low-magnetic field WAL signals are not well resolved, which strongly indicates that the contribution by a 3D conductivity in the bulk hampers its observation. However, in micro flakes the 3D bulk contribution to the conductivity is strongly reduced as the surface-to-volume ratio is increased by a factor of up to 1000. We clearly observe the WAL conductivity cusps around zero magnetic field. For metallic-like micro flakes the low-temperature electron density is about that of the bulk (macro flakes), but the mobility can be increased, e.g. for 130~nm thin flake $n_\mathrm{3D}\!=\!1.2\cdot10^{19}$cm$^{-3}$ and $\mu_\mathrm{3D}\!=\!2320$ cm$^2$/(Vs). In the semiconducting case we find a strongly reduced density but a bulk-like mobility, e.g. for a 220~nm thick flake $n_\mathrm{3D}\!=\!1.2\cdot10^{17}$cm$^{-3}$ and $\mu_\mathrm{3D}\!=\!675$ cm$^2$/(Vs). In the metallic-like case a higher density ($10^{19}$ cm$^{-3}$ vs. $10^{17}$ cm$^{-3}$) may lead to two effects which enhance the mobility: first, a screening of potential fluctuations and therefore enhanced scattering times, and second, an increase of the Fermi level and hence additional kinetic energy. In order to identify contributions of the TSS to the transport, we studied the change of the magnetoconductivity $\Delta\sigma_\mathrm{xx}$ in perpendicular low-magnetic fields as shown in Fig.~\ref{fig6}(a) and (b). A WAL maximum is clearly visible near zero magnetic field and is suppressed for higher temperatures. At higher magnetic fields the total magnetoresistance has additional contributions $\propto\!B^2$. 

Typically, the low-field behavior is associated with WAL originating from either strong spin-orbit interaction in the bulk and/or spin-momentum locking in the topological surface states.\cite{Ando-2013-JPSJ,Altshuler,Matsuo} In thin Bi$_2$Se$_3$ films or micro flakes the assumption usually is made that these can be considered as 2D systems with strong spin-orbit interaction, and the HLN model\cite{hik80} is used to analyze the 2D magnetoconductivity. Such a model can be applied as long as the dephasing time $\tau_\phi$ is much smaller than the spin-orbit time $\tau_{\mathrm{SO}}$ and the inelastic scattering (energy relaxation) time $\tau_{\mathrm{e}}$, i.e. $\tau_\phi\!\ll\!\tau_{\mathrm{SO}}$ and $\tau_\phi\!\ll\!\tau_{\mathrm{e}}$. We applied the HLN model to the temperature-dependent magnetoconductance data plotted in Fig.~\ref{fig5}(a) and (b) by considering: 
\begin{eqnarray}
\Delta\sigma_\mathrm{xx}(B)=\sigma_\mathrm{xx}(B)-\sigma_\mathrm{xx}(B\!=\!0)=\alpha\frac{e^{2}}{2\pi^{2}\hslash}\left[\psi\left(\frac{1}{2}+\frac{B_\phi}{B}\right)-\ln\left(\frac{B_\phi}{B}\right)\right]\,,
\end{eqnarray}
where $\psi(x)\!=\!\Gamma^{\prime}(x)/\Gamma(x)\!=\!\mathrm{d}\ln\Gamma(x)/\mathrm{d}x$ represents the digamma function. The prefactor $\alpha$ can be used to estimate the number of independent channels contributing to the interference: $\alpha\!=\!-0.5$ for a single coherent topological surface channel contributing to the WAL cusp, and $\alpha\!=\!-1$ for two independent coherent transport channels.\cite{lan13} $B_\phi\!=\!\hslash/(4e\ell_\phi^2)$ is the characteristic magnetic field and $\ell_\phi$ is the phase coherence length. In the analysis $\alpha$ and $\ell_\phi$ are fitting parameters. The symbols in Fig.~\ref{fig6}(a) and (b) represent experimental data and the curves correspond to fits to the HLN model (see Eq. (1)). 

This analysis can be performed in a straightforward manner in the case of the semiconducting flake (see Fig.~\ref{fig6}b). As TSS are directly visible in the ARPES spectra, our evaluation is in agreement with the conventional interpretation of TSS contributions to WAL in Bi$_2$Se$_3$. However, this analysis cannot be applied directly to the conductivity data of the metallic-like micro flakes (see Fig.~\ref{fig6}a). In these, the change of the magnetoconductivity with the applied perpendicular magnetic field is by a factor of up to 100 too large, so that $\alpha$ cannot be obtained in a valid range within the HLN model. Due to the QHE observed in the metallic-like macro flakes, we identify the observed low-magnetic field conductivity as the result of 2D layered transport. Therefore, the HLN fit is applied to the magnetoconductivity divided by the number
$Z^*$ of contributing 2D layers, which is about half the number of QLs, i.e. $Z^*\!=\!0.5\times t/(\mathrm{1~nm})$. Therefore, we define a magnetoconductivity per 2D layer: $\Delta\widetilde{\sigma}_\mathrm{xx}\!=\!\Delta\sigma_\mathrm{xx}/Z^*$ which is shown in Fig.~\ref{fig6}(a).

In Fig.~\ref{fig6}(c) and (d), $\alpha$ and $\ell_\phi$ of the HLN fits for two metallic-like and one semiconducting case at different temperatures are shown. At the lowest temperature of 300~mK the $\alpha$-values for the semiconducting ($t\!=\!220$ nm) and thin ($t\!=\!130$ nm) metallic-like micro flakes are about $-$0.5 and decrease in magnitude with increasing temperature, which is in accordance with previous reports.\cite{tas12,ste11,che11} This indicates that in the semiconducting case two coupled surface states dominate the WAL behavior in accordance with the HLN model. Furthermore, in the metallic-like case the assumption of $Z^*$ 2D layers allows a fit by the HLN model. For the thicker metallic-like case $\alpha\!=\!-0.25$ is obtained in the above manner. This lowered value could indicate more contributions from the bulk. The application of the HLN model appears feasible under the assumption that mainly $Z^*$ distinct 2D layers contribute equally - which would resemble the case of parallel conducting layers. This is valid if the 2D layer contribution dominates the WAL cusp.

The values of $\ell_\phi$ remain largely unaffected by such interpretations and are given in Fig.~\ref{fig6}(d). At 300~mK $\ell_\phi$ is in agreement with the flake thickness of the metallic-like cases, which in general indicates that an assumption of a 2D system may be applied. For the semiconducting case $\ell_\phi$ is with 40~nm about one fifth of the flake thickness which may indicate an effective depth of about 20~nm in which the topological surface states extend into the bulk. 

In conclusion, from application of the HLN model to the low-temperature and low-magnetic field conductivity we find that for a semiconducting flake the topological surface states can be identified from the WAL, as expected from our ARPES measurements. However, in the metallic-like micro flakes the strong decrease in magnetoconductivity with a magnetic field finds its explanation in the HLN model only if effectively more 2D layers than the two surface layers contribute. Combining the ARPES and QHE results from the macro flakes, we find that the simple HLN model can be applied if $Z^*$ coupled 2D layers contribute. In general, band bending at the surface could lead to additional 2D layers, however, we found no indications for band bending by different mobilities from Hall measurements and signatures in ARPES. Instead, the indications from the TEM analysis on the structural stacking disorder along the $c$-axis direction may be indicative for the origin of additional 2D layers in our transport experiments.

\section*{Discussion}

In summary, TSS as detected from ARPES measurements on Bi$_2$Se$_3$ single crystals are confirmed by transport measurements on semiconducting micro flakes. TEM and PEEM analysis ensures that micro flakes have the same structural and chemical composition as in the bulk. Additionally, our study reveals uniquely metallic-like behavior, which shows pronounced 2D effects in the {\it high- and low-field} magnetotransport. Most probably this occurs by contact to ambient atmospheres (air) and annealing steps during the micro contact preparation (see section Methods), leading to selenium vacancies in the near surface region. The comprehensive experiments suggest that 2D layered transport plays a decisive role in highly conductive Bi$_2$Se$_3$. Therefore, contributions to the magnetoresistance cannot be simply classified as originating only from the TSS or 3D bulk. Instead additional 2D layered transport {\it in the bulk} are important in $n$-type Bi$_2$Se$_3$ in the high electron-density regime \mbox{($\sim\!1.2\cdot10^{19}-1.9\cdot10^{19}$ cm$^{-3}$)}, which behaves in a more complex manner than has been stated before. Both TSS at the surface and 2D layered transport in the bulk contribute to the localization phenomena. Based on our high- and low-magnetic field data we conclude that the 2D layers need not necessarily be homogenously distributed across the bulk cross section, but may contribute mostly from the near-surface regions. One reason may be the effective thickness of a Selen-depleted surface region, which hosts the additional 2D transport layers. The HLN fit can be successfully performed under the assumption of $Z^*$ stacked 2D conducting layers. This clearly indicates that a significant number of conducting 2D layers contribute, in the order of the number of quintuple layers in the micro flake (i.e. $\sim\!100$). This is much more than two possible TSS states alone. Therefore, we conclude that additionally to the (maximum) two TSS - and possibly other 2D channels due to band bending (maximum a few), there are up to the number of quintuple layers 2D channels which contribute. Our transport results in correlation with the ARPES experiments prove the coexistence of TSS and 2D layered transport. This may lead to a broad range of novel phenomena in highly conductive 3D topological insulator materials.

\section*{Methods}

\subsection*{Sample preparation and characterization}

High-quality single crystalline Bi$_2$Se$_3$ were prepared from melt with the Bridgman technique.\cite{sht09} The growth time, including cooling was about 2 weeks for a $\sim$50 g crystal. The carrier density of the resulting samples was about $1.9\cdot10^{19}$ cm$^{-3}$, as determined by Hall measurements. The whole crystal was easily cleaved along the [00.1] growth direction, indicating crystal perfection. Macro and micro flakes were prepared by cleaving the single crystal. Macro flakes were prepared with a thickness of around 110 $\mu$m to investigate bulk properties. The micro flakes with a thickness in a range from 40 nm up to 300 nm were prepared using a mechanical exfoliation technique similar to the one used for graphene. The exfoliation procedure was carried out on top of a 300 nm thick SiO$_2$ layer grown on a boron-doped Si substrate. The preparation of the micro flakes involved in general the following steps: (i) Prior to sample preparation, the substrates were cleaned with acetone in an ultrasonic bath for about 3$-$4 minutes. After sonication in acetone, drops of ethanol were placed on the substrates and subsequently blow-dried with nitrogen. (ii) A $5\!\times\!5\!\times\!1$ mm$^3$ piece of the initial Bi$_2$Se$_3$ crystal was glued on a separate Si substrate with a GE7031 varnish. Using adhesive tape, a thin layer of Bi$_2$Se$_3$ was cleaved and then folded and unfolded back several times into the adhesive tape, resulting in a subsequent cleaving of the layer into thinner and thinner flakes. (iii) The SiO$_2/$Si substrate was then placed on top of the adhesive tape in a region uniformly covered by Bi$_2$Se$_3$ micro flakes. By pressing gently on top of the substrate with tweezers, the Bi$_{2}$Se$_{3}$ micro flakes were placed on the surface of the Si wafer due to the applied mechanical stress. In Fig. S2 in the supplemental information we show an atomic force microscopy (AFM) image of a selected Bi$_2$Se$_3$ flake homogeneous in thickness, drop-cast on the SiO$_2$ substrate. While the metallic-like macro flakes were prepared at room temperature without any annealing, during the micro contact preparation for the metallic-like micro flakes several annealing steps with photoresist were applied. For the semiconducting micro flakes the annealing was reduced to 1 minute at $\sim100\,^{\circ}{\rm C}$. 

We explore the structural properties of the flakes with AFM, STEM and HRTEM. The flake composition and surface stability are investigated using energy-dispersive x-ray spectroscopy (EDX), see supplemental information, and spatially resolved core-level X-ray PEEM. Structural analysis using HRTEM and STEM was carried out at a JEOL JEM2200FS microscope operated at 200~kV. The sample preparation for HRTEM characterization consisted of ultrasonic separation of the flakes from the substrate, followed by their transfer onto a carbon-coated copper grid. Using adhesive tape, the surface was prepared by cleavage of the crystal along its trigonal axis in the direction perpendicular to the van-der-Waals-type $(0001)$ planes. ARPES measurements were performed at a temperature of 12~K in UHV at a pressure of $\sim5\cdot10^{-10}$ mbar with a VG Scienta R8000 electron analyzer at the UE112-PGM2a beamline of BESSY II using p-polarized undulator radiation. The micro flakes were characterized by spatially-resolved core-level microspectroscopy using an Elmitec PEEM instrument at the UE49-PGMa microfocus beamline of BESSY II.

After the micro flake selection micro-laser lithography was used to fabricate the contacts for the transport measurements. Positive photoresist AZ ECI 3027 was coated, and the samples were spun in air for 45~s at speeds varying from 2500 to 3000~rpm. After micro-laser lithography Ti/Au (10~nm/40~nm) were sputtered and lift-off processing was used to prepare the contacts. Semiconducting behavior is only observed in micro flakes if contact to air is avoided and heating procedures are minimized during the device processing. For macro flakes contacts were prepared with Ag paint and Au wires. All samples were stored in dry N$_2$ atmosphere.

Using a He-3 cryostat, temperature-dependent four-terminal resistance measurements were performed in a range from 300~K down to 0.3~K. The magnetoresistivities $\rho_\mathrm{xx}$ and $\rho_\mathrm{xy}$ were measured in a temperature range from 0.3~K up to several K. The macro flakes were measured with the Van der Pauw method\cite{Vanderpauw-1958-PRR}, using a \textit{Keithley} 6221 current source and a \textit{Keithley} 2182A nanovoltmeter in
"Delta mode" with a current $|I|\!=\!100~\mu$A. The micro flakes were measured with Hall-bar configurations, using either DC techniques (i.e., using a \textit{Keithley} 6221 current source and a \textit{Keithley} 2182A nanovoltmeter) with currents up to $|I|\!=\!1~\mu$A or low-frequency AC techniques (i.e. using \textit{Signal Recovery} 7265 lock-in amplifiers) with currents $|I|$ between 2~nA and 20~nA.

\section*{Acknowledgements}
Financial support from the Deutsche Forschungsgemeinschaft within the priority program SPP1666 (Grant No. FI932/7-1 and RA1041/7-1) and the Bundesministerium f\"ur Bildung und Forschung (Grant No. 05K10WMA) is gratefully acknowledged.

\section*{Author contributions statement}
L.V.Y. conducted the bulk crystal growth, O.C., C.R., D.L., M.B., S.D., and S.F.F. contributed to structural characterization and transport experiments, A.M. and S.D. performed the HRTEM, STEM and EDX analysis, J.S.-B., S.V., A.A.\"U., and O.R. conducted ARPES and PEEM experiments, O.C., O.R., J.S.-B., M.B., and S.F.F. conceived the experiments, analyzed the data and wrote the manuscript. All authors contributed to the discussion and reviewed the manuscript. 

\section*{Additional information}
\textbf{Competing financial interests:} The authors declare no competing financial interests. 
\end{document}